%% file: barbosa.tex
\documentclass{edbk}
\usepackage{edbkps}
\usepackage{amsfonts}
\usepackage{epsfig}
\usepackage[latin1]{inputenc}
\input{def}

\draft
\normallatexbib
\begin{document}

\def\inxx{}

\articletitle[The Combinatorics of Resource Sharing]{The Combinatorics of\\ Resource Sharing}

\author{Valmir C. Barbosa}
\affil{Universidade Federal do Rio de Janeiro\footnote{Programa de Engenharia
de Sistemas e Computa\c c\~ao, COPPE, Caixa Postal 68511, 21945-970 Rio
de Janeiro - RJ, Brazil. This author is supported by the Brazilian agencies
CNPq and CAPES, the PRONEX initiative of Brazil's MCT under contract
41.96.0857.00, and by a FAPERJ BBP grant.}}
\email{valmir@cos.ufrj.br}


\begin{abstract}
We discuss general models of resource-sharing computations, with emphasis on
the combinatorial structures and concepts that underlie the various deadlock
models that have been proposed, the design of algorithms and deadlock-handling
policies, and concurrency issues. These structures are mostly graph-theoretic
in nature, or partially ordered sets for the establishment of priorities among
processes and acquisition orders on resources. We also discuss graph-coloring
concepts as they relate to resource sharing.
\end{abstract}

\def\Kappa{\rm K}

\section{Introduction}\label{intr}

The\inxx{resource sharing}
sharing of resources by processes under the requirement of mutual exclusion
is one of the most fundamental issues in the design of computer systems, and
stands at the crux of most efficiency considerations for those systems. When
referred to with such generality, processes can stand for any of the computing
entities one finds at the various levels of a computer system, and likewise
resources are any of the means necessary for those entities to function.
Resources tend to be scarce (or to get scarce shortly after being made
available), so the designer of a computer system at any level must get involved
with the task of devising allocation policies whereby the granting of resources
to processes can take place with at least a minimal set of guarantees.

One such guarantee is of the so-called
{\em safety\/}\inxx{resource sharing, safety}
type, and in essence forbids the occurrence of
{\em deadlock\/}\inxx{resource sharing, deadlock} situations.
A deadlock situation is characterized by the permanent impossibility for a
group of processes to progress with their tasks due to the occurrence of a
condition that prevents at least one needed resource from being granted to each
of the processes in that group. Another guarantee one normally seeks is a
{\em liveness\/}\inxx{resource sharing, liveness} guarantee,
which imposes bounds on the wait that any process
must undergo between requesting and being granted access to a resource, and
thereby ensures that {\em lockout\/}\inxx{resource sharing, lockout}
situations never happen.

There are difficulties of various sorts associated with designing and analyzing
resource-sharing policies. Some of these difficulties refer to the choice and
use of mathematical models that can account properly for the relevant details
of the resource-sharing problem at hand. Similarly, there are difficulties that
stem from the inherent asynchronism that typically characterizes the behavior
of processes in a computer system. This asynchronism, though essential in
depicting most computer systems realistically, tends to introduce subtle
obstacles to the design of correct algorithms.

In this paper, we are concerned with the several combinatorial models that
have proven instrumental in the design and analysis of resource-sharing
policies. The models that we consider are essentially of graph-theoretic nature,
and relate closely to the aforementioned safety and liveness issues. The
essential notation that we use is the following. The set of processes is
denoted by ${\cal P}=\{P_1,\ldots,P_n\}$, and the set of resources by
${\cal R}=\{R_1,\ldots,R_m\}$. For $P_i\in{\cal P}$,
${\cal R}_i\subseteq{\cal R}$ is the set of resources to which $P_i$ may request
access. Similarly, for $R_p\in{\cal R}$, ${\cal P}_p\subseteq{\cal P}$ is the
set of processes that may request access to $R_p$. Clearly, for $1\le i\le n$
and $1\le p\le m$, $P_i\in{\cal P}_p$ if and only if $R_p\in{\cal R}_i$. Also,
we let ${\cal R}_{ij}={\cal R}_i\cap{\cal R}_j$ for $1\le i,j\le n$,
and ${\cal P}_{pq}={\cal P}_p\cap{\cal P}_q$ for $1\le p,q\le m$.

\begin{exmp}\label{basic}
If ${\cal P}=\{P_1,P_2,P_3,P_4,P_5\}$ and
${\cal R}=\{R_1,R_2,R_3,R_4,R_5,R_6\}$
with ${\cal R}_1=\{R_1,R_2\}$, ${\cal R}_2=\{R_2,R_3,R_6\}$,
${\cal R}_3=\{R_3,R_4,R_6\}$, ${\cal R}_4=\{R_4,R_5,R_6\}$, and
${\cal R}_5=\{R_1,R_5\}$, then ${\cal P}_1=\{P_1,P_5\}$,
${\cal P}_2=\{P_1,P_2\}$, ${\cal P}_3=\{P_2,P_3\}$, ${\cal P}_4=\{P_3,P_4\}$,
${\cal P}_5=\{P_4,P_5\}$, and ${\cal P}_6=\{P_2,P_3,P_4\}$. In addition, we
have the nonempty sets shown in Table~\ref{subsets}.
\end{exmp}

\begin{table}[ht]
\caption{Resource and process sets for Example~\ref{basic}\label{subsets}.}
\centerline{\epsfbox{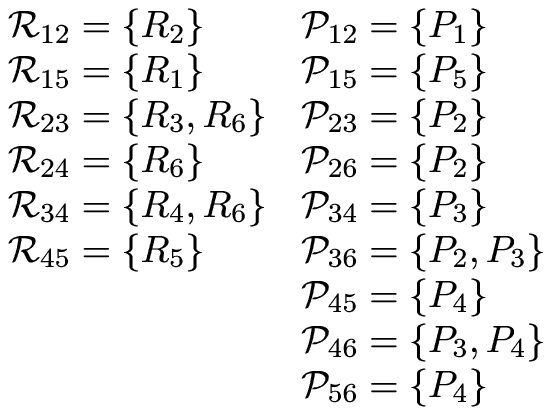}}
\end{table}

The following is how the remainder of the paper is organized.
In Section~\ref{comp}, we provide an outline of the generic computation that
is carried out by the members of ${\cal P}$ in order to share the resources
in ${\cal R}$. Such an outline is given as an asynchronous distributed
algorithm, and aims at emphasizing the communication that must take place
among processes for resource sharing. This communication comprises at least
messages for requesting and granting access to resources. Depending on how
such messages are composed and handled by the processes, one gets one of the
various {\em deadlock models\/} that have appeared in the literature. These
models are our subject in Section~\ref{models}. The two sections that follow
(Sections~\ref{detect} and~\ref{prevent}) are devoted to the combinatorics
underlying the two broad classes of deadlock-handling policies, namely those
of {\em detection\/} and {\em prevention\/} strategies, respectively.
We then move,
in Section~\ref{abacus}, to a prevention policy that generalizes one of policies
discussed in Section~\ref{prevent} and for which an abacus-like graph structure
is instrumental. This generalized policy is for the case of high demand for
resources by the processes. Section~\ref{color} discusses the relationship that
exists between concurrency in resource sharing and the various chromatic
indicators of a graph. Concluding remarks follow in Section~\ref{concl}.

In this paper, all lemma and theorem proofs are omitted, but references are
given to where they can be found.

\section{Resource-sharing computations}\label{comp}

\begin{figure}[ht]
\centerline{\epsfbox{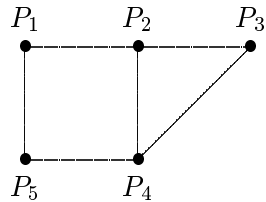}}
\caption{The graph $G$ for Example~\ref{basic}\label{graphg}.}
\end{figure}

The\inxx{resource-sharing computation}
model of computation that we assume in this section is the standard
{\em fully asynchronous\/} (or simply
{\em asynchronous\/})\inxx{asynchronous model of distributed computing}
model of distributed computing \cite{b96}.
In this model, every member of ${\cal P}$ possesses a local,
independent clock, having therefore a time basis that is totally uncorrelated
to that of any other process. In addition, all communication among processes
take place via point-to-point message passing, requiring a finite (though
unpredictable) time for message delivery. Messages are sent over bidirectional
communication channels, of which there exists one for every $P_i,P_j\in{\cal P}$
such that ${\cal R}_{ij}\neq\emptyset$. That is, every two processes with the
potential to share at least one resource are directly interconnected by a
bidirectional communication channel. If we let ${\cal C}$ denote the set of
such channels, then the undirected graph $G=({\cal P},{\cal C})$, having one
vertex for each process and one edge for each channel, can be used to represent
the system over which our resource-sharing computations run. In $G$, and for
$1\le p\le m$, the vertices in ${\cal P}_p$ induce a completely connected
subgraph (a {\em clique\/}\inxx{undirected graph, clique}
\cite{bm76}).
We assume that $G$ is a connected graph, as processes
belonging to different connected components never interfere with each other.
In the context of Example~\ref{basic}, $G$ is the graph shown in
Figure~\ref{graphg}.

In the computations that we consider, a process executes the following four
procedures.

\begin{itemize}
\item {\sc Request};
\item {\sc Check\_Priority};
\item {\sc Compute};
\item {\sc Clean\_up}.
\end{itemize}

Each of these procedures is executed atomically in response to a specific
event, as follows. When the need arises for the process to compute on shared
resources, it executes the {\sc Request} procedure. Typically, this will entail
sending to some of its neighbors in $G$ messages requesting exclusive access
to resources shared with them. The reception of one such message causes the
receiving process to execute {\sc Check\_Priority}, whose outcome will guide
the process' decision as to whether grant or not the requested exclusive
access. If the process does decide to grant the request, then a message
carrying this information is sent back to the requesting process, which upon
receipt executes {\sc Compute}. This procedure is a test to see whether
the process already holds exclusive access to enough resources to carry out
its computation, which it does in the affirmative case; it keeps waiting,
otherwise. If and when the resource-sharing computation is completed, the
process engages in a message exchange with its neighbors in $G$ by executing
the {\sc Clean\_up} procedure. This message exchange may revise priorities
and cause previously withheld requests to be granted.

This outline is admittedly far too generic in several aspects, but already it
provides the background for the key questions underlying the establishment of
a resource-sharing policy. For example: To which resources does a process
request exclusive access in {\sc Request} when in need for shared resources? 
At which point when executing {\sc Compute} does it decide it may proceed with
its computation? How do the {\sc Check\_Priority} and {\sc Clean\_up} procedures
cooperate to handle the priority issue properly? Answers to these questions have
been given
in the context of several application areas, and along with numerous models and
algorithms. Addressing them in detail is beyond our intended scope, but in
Section~\ref{models} we present the abstraction of deadlock models, which
summarizes the issues that are critical to our discussion of the combinatorics
of resource sharing.

Note that both the execution of {\sc Request} and the initial test performed by
{\sc Compute} may entail waiting on the part of the calling process. Clearly,
then, and depending on how the priority issue is handled, here lies the
possibility for unbounded wait, which is directly related to the safety and
liveness guarantees we may wish to provide. The approaches here vary greatly,
and may be grouped into two broad categories. On the ``optimistic'' side, one
may opt for a somewhat loose priority scheme and risk the loss of those
guarantees. In such cases, the loss of safety leads to the need for the
capability of detecting deadlocks \cite{cmh83,k87,s89}. The opposing, more
``conservative'' side is the side of those strategies which ``by design''
guarantee safety and liveness, thereby preventing their loss beforehand.

As we demonstrate in the remainder of the paper, both categories give rise to
interesting combinatorial structures and properties, especially as they relate
to the deadlock issue. We then end this section by defining what will be meant
henceforth by deadlock, although still somewhat informally. As we go through
the various combinatorial structures that relate closely to deadlocks, such
informality will dissipate. A subset of processes ${\cal S}\subseteq{\cal P}$
is in deadlock\inxx{resource sharing, deadlock}
if and only if every process in ${\cal S}$ is waiting for a
condition that ultimately can be relieved only by another member of ${\cal S}$
whose own wait is over. Obviously, then, deadlocks are stable properties: Once
they take hold of a group of processes, only the external intervention that
eventually follows detection may break them. Prevention strategies, by contrast,
seek never to let them happen.

\section{Deadlock models}\label{models}

A\inxx{deadlock model}
deadlock model is an abstraction of the rules that govern the wait of
processes for one another as they execute the procedures {\sc Request},
{\sc Check\_Pri\-ority}, {\sc Compute}, and {\sc Clean\_up} discussed in
Section~\ref{comp}. Deadlock models are defined on top of a dynamic graph,
called the
{\em wait-for graph\/}\inxx{resource-sharing computation, wait-for graph}
and henceforth denoted by $W$.

$W$ is the directed graph $W=({\cal P},{\cal W})$, having the same vertex set
as $G$ (one vertex per process) and the directed edges in ${\cal W}$. This set
is such that an edge exists directed from process $P_i$ to process $P_j$ if
and only if $P_i$ has sent $P_j$ a request for exclusive access to some
resource that they share and is waiting either for $P_j$ to grant the request
or for the need for that resource to cease existing as grant messages are
received from other processes. For $P_i\in{\cal P}$, we let
${\cal O}_i\subseteq{\cal P}$ be the set of processes towards which edges are
directed away from $P_i$ in ${\cal W}$.

It follows from the definition of $W$ that the only processes that may be
carrying out some computation on shared resources are those that are
{\em sinks\/}\inxx{directed graph, sink}
in $W$ (vertices with no adjacent edges directed outward,
including isolated vertices). All other processes are waiting for exclusive
access to the resources they need. Clearly, then, a
necessary\inxx{resource-sharing computation, necessary condition for deadlock}
condition for a
deadlock to exist in $W$ is that $W$ contain a directed cycle.

\begin{fact}\label{cycle}
If a deadlock exists in $W$, then $W$ contains a directed cycle.
\end{fact}

\begin{figure}[ht]
\centerline{\epsfbox{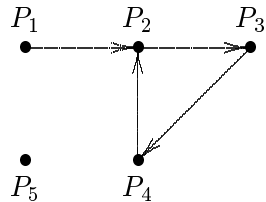}}
\caption{The graph $W$ for Example~\ref{deadlock}\label{graphw}.}
\end{figure}

\begin{exmp}\label{deadlock}
In the context of Example~\ref{basic}, suppose a deadlock has happened involving
processes $P_2$, $P_3$, and $P_4$. Suppose also that process $P_1$ is waiting
for resource $R_2$, which is held by $P_2$, which in turn is waiting for $R_3$,
held by $P_3$, which is waiting for $R_4$, held by $P_4$. If, in addition, $R_6$
is held by $P_2$ and awaited by $P_4$, then the corresponding $W$ is the one
shown in Figure~\ref{graphw}, with the directed cycle on $P_2$, $P_3$, and
$P_4$.
\end{exmp}

In Section~\ref{detect}, after we have gone through a variety of deadlock
models in the remainder of this section, we will come to the conditions that
are sufficient for a deadlock to exist in $W$.

This is the sense in which graph $W$ is a dynamic structure: Although its vertex
set is always the same, as the processes interact with one another by executing
the aforementioned procedures, its edge set changes. Normally, given the view
we have adopted of the resource-sharing computation as an asynchronous
distributed computation, one must bear in mind the fact that it only makes
sense to refer to $W$ as associated with some
{\em consistent\inxx{consistent global state}
global state\/}
of the computation \cite{b96,cl85}.
For our purposes, however, such an association does not
have to be explicit, so long as one understands the dynamic character of $W$.

What determines the evolution of $W$ by allowing for changes in the set
${\cal W}$ of directed edges is the deadlock model that holds for the
computation. The deadlock models that have been investigated to date are the
ones we discuss next. In essence, what each of these deadlock models does is
to specify rules for vertices that are not sinks in $W$ to become sinks.

\paragraph{The AND model}\inxx{deadlock model, AND model}
In the AND model, a process $P_i$ can only become a
sink when its wait is relieved by all processes in ${\cal O}_i$. This model
characterizes, for example, situations in which a conjunction of resources is
needed by $P_i$ \cite{bhrs95,ks94,rp95}.

\paragraph{The OR model}\inxx{deadlock model, OR model}
In the OR model, it suffices for process $P_i$ to
be relieved by one of the processes in ${\cal O}_i$ in order for its wait to
finish. The OR model characterizes, for example, some of the situations in
which any one of a group of resources (a disjunction of resources) is needed by
$P_i$ \cite{bhrs95,ks94,mc82,rp95}.

\paragraph{The $x$-out-of-$y$ model}\inxx{deadlock model, $x$-out-of-$y$ model}
In this model, there are two integers,
$x_i$ and $y_i$, associated with process $P_i$. Also,
$y_i=\vert{\cal O}_i\vert$, meaning that process $P_i$ is in principle waiting
for communication from every process in ${\cal O}_i$. However, in order to be
relieved from its wait condition, it suffices that such communication arrive
from any $x_i$ of those $y_i$ processes. The $x$-out-of-$y$ model can then be
used, for example, to characterize situations in which $P_i$ starts by requiring
access permissions in excess of what it really needs, and then withdraws the
requests that may still be pending when the first $x_i$ responses are received
\cite{bt87,bhrs95,ks94}.

\paragraph{The AND-OR model}\inxx{deadlock model, AND-OR model}
In the AND-OR model, there are $t_i\ge 1$ subsets
of ${\cal O}_i$ associated with process $P_i$. These subsets are denoted by
${\cal O}_i^1,\ldots,{\cal O}_i^{t_i}$ and must be such that
${\cal O}_i={\cal O}_i^1\cup\cdots\cup{\cal O}_i^{t_i}$.
In order for process $P_i$ to be relieved from its wait condition, it must
receive grant messages from all the processes in at least one of
${\cal O}_i^1,\ldots,{\cal O}_i^{t_i}$. For this reason, these $t_i$ subsets of
${\cal O}_i$ are assumed to be such that no one is contained in another.
Situations that the AND-OR model characterizes are, for example, those in which
$P_i$ perceives several conjunctions of resources as equivalent to one another
and issues requests for several of them with provisions to withdraw some of them
later \cite{bb98,bhrs95,ks94,rp95}.

\paragraph{The disjunctive $x$-out-of-$y$
model}\inxx{deadlock model, disjunctive $x$-out-of-$y$ model}
In this model, associated
with process $P_i$ are $u_i\ge 1$ pairs of integers, denoted by
$(x_i^1,y_i^1),\ldots,(x_i^{u_i},y_i^{u_i})$. These integers are such that
$y_i^1=\vert{\cal Q}_i^1\vert,\ldots,y_i^{u_i}=\vert{\cal Q}_i^{u_i}\vert$,
where ${\cal Q}_i^1,\ldots,{\cal Q}_i^{u_i}$ are subsets of
${\cal O}_i$ such that ${\cal O}_i={\cal Q}_i^1\cup\cdots\cup{\cal Q}_i^{u_i}$.
In order to be relieved from its wait condition, $P_i$ must be granted access
to shared resources by either $x_i^1$ of the $y_i^1$ processes in
${\cal Q}_i^1$, or $x_i^2$ of the $y_i^2$ processes in ${\cal Q}_i^2$, and so
on. Of course, it makes no sense for
${\cal Q}'_i,{\cal Q}''_i\in\{{\cal Q}_i^1,\ldots,{\cal Q}_i^{u_i}\}$ to exist
such that ${\cal Q}'_i\subseteq{\cal Q}''_i$ and $x'_i\ge x''_i$, which is then
assumed not to be the case. This model characterizes situations similar to those
characterized by the $x$-out-of-$y$ model, and generalizes that model by
allowing for a disjunction on top of it
\cite{bhrs95,ks94}.

\vskip12pt
As one readily realizes, these five models are not totally uncorrelated and
a strict hierarchy exists in which a model generalizes the previous one in the
sense that it contains as special cases all the possible wait conditions of the
other. For example, the $x$-out-of-$y$ model generalizes the AND model with
$x_i=y_i$ and the OR model with $x_i=1$ for all $P_i\in{\cal P}$. Likewise,
and also for all $P_i\in{\cal P}$, the AND-OR model also generalizes the AND
model with $t_i=1$ and the OR model with
$\vert{\cal O}_i^1\vert=\cdots=\vert{\cal O}_i^{t_i}\vert=1$.

Despite this ability of both the $x$-out-of-$y$ model and the AND-OR model
to generalize both the AND and OR models, they are not equivalent to each
other. In fact, the AND-OR model is more general than the $x$-out-of-$y$ model,
while the converse is not true. In order for the AND-OR model to
express a general $x$-out-of-$y$ condition, it suffices that, for all
$P_i\in{\cal P}$, $t_i={{y_i}\choose{x_i}}$ and
$\vert{\cal O}_i^1\vert=\cdots=\vert{\cal O}_i^{t_i}\vert=x_i$.

\begin{exmp}\label{xyandor}
Suppose that we have, for some $P_i\in{\cal P}$,
${\cal O}_i=\{P_j,P_k,P_\ell\}$. In the $x$-out-of-$y$ model, $y_i=3$. If
$x_i=2$, then in the AND-OR model we have, equivalently,
$t_i={{3}\choose{2}}=3$,
${\cal O}_i^1=\{P_j,P_k\}$, ${\cal O}_i^2=\{P_j,P_\ell\}$, and
${\cal O}_i^3=\{P_k,P_\ell\}$.
\end{exmp}

To finalize our discussion on how the five deadlock models are related to
one another, note that the AND-OR model and the disjunctive $x$-out-of-$y$
model are equivalent to each other. In order to see that the AND-OR model
generalizes the disjunctive $x$-out-of-$y$ model, let
$t_i=v_i^1+\cdots+v_i^{u_i}$, where
\[1\le v_i^1\le{{y_i^1}\choose{x_i^1}},\ldots,
1\le v_i^{u_i}\le{{y_i^{u_i}}\choose{x_i^{u_i}}},\]
for all $P_i\in{\cal P}$. In addition, $v_i^1$ of the sets
${\cal O}_i^1,\ldots,{\cal O}_i^{t_i}$ must have cardinality $x_i^1$ and be
subsets of ${\cal Q}_i^1$, the same holding for the other superscripts
$2,\ldots,u_i$, and explicit care must be exercised to avoid any of the sets
${\cal O}_i^1,\ldots,{\cal O}_i^{t_i}$ being a subset of another.

That the disjunctive $x$-out-of-$y$ model generalizes the AND-OR model is
simpler to see. For such, it suffices that, for all $P_i\in{\cal P}$, we let
$u_i=t_i$ and
${\cal Q}_i^1={\cal O}_i^1,\ldots,{\cal Q}_i^{u_i}={\cal O}_i^{t_i}$,
along with $x_i^1=y_i^1,\ldots,x_i^{u_i}=y_i^{u_i}$.

\begin{exmp}\label{dxyandor}
Let ${\cal O}_i=\{P_j,P_k,P_\ell,P_t\}$ for some $P_i\in{\cal P}$.
In the disjunctive $x$-out-of-$y$ model, suppose we have $u_i=2$,
${\cal Q}_i^1=\{P_j,P_k\}$, and ${\cal Q}_i^2=\{P_k,P_\ell,P_t\}$,
yielding $y_i^1=2$ and $y_i^2=3$. If $x_i^1=x_i^2=2$, then in the AND-OR model
we have $t_i={{2}\choose{2}}+{{3}\choose{2}}=4$,
${\cal O}_i^1=\{P_j,P_k\}$, ${\cal O}_i^2=\{P_k,P_\ell\}$,
${\cal O}_i^3=\{P_k,P_t\}$, and ${\cal O}_i^4=\{P_\ell,P_t\}$.
Had we started out with this AND-OR setting, then for the disjunctive
$x$-out-of-$y$ model we would have ${\cal Q}_i^1=\{P_j,P_k\}$,
${\cal Q}_i^2=\{P_k,P_\ell\}$, ${\cal Q}_i^3=\{P_k,P_t\}$, and
${\cal Q}_i^4=\{P_\ell,P_t\}$. We would also have
$x_i^1=y_i^1=x_i^2=y_i^2=x_i^3=y_i^3=x_i^4=y_i^4=2$.
Clearly, this is equivalent to the disjunctive $x$-out-of-$y$ scenario of the
beginning of this example.
\end{exmp}

\section[Graph structures for deadlock detection]{Graph structures for\\ deadlock detection}\label{detect}

As\inxx{deadlock detection}\inxx{resource-sharing computation, sufficient condition for deadlock}
we remarked in Section~\ref{comp}, computations that make no {\em a priori\/}
provisions against the occurrence of deadlocks must, if the need arises, resort
to techniques for the detection of deadlocks. Detecting the existence of a
deadlock in the wait-for graph $W$ can become the detection of a graph-theoretic
property on $W$ if we are able to characterize conditions on $W$ that are
sufficient for the existence of deadlocks. As we discuss in this section, such
a property exists for all the deadlock models of Section~\ref{models}. However,
not always is it the case that detecting this graph-theoretic property directly
is the most efficient means of deadlock detection. When this
happens not to be the
case, alternative approaches must be employed, usually based on some form of
simulation of the sending of grant messages.

We\inxx{deadlock detection, AND model}
start with the AND model, and recognize immediately that the presence of
a directed cycle in $W$ is not only a necessary condition for the existence of
a deadlock in $W$ but also a sufficient condition. This is so because, in the
AND model, every process requires grant messages to be received on all edges
directed away from it, which clearly is precluded by the existence of a directed
cycle.

\begin{fact}\label{andsuff}
In the AND model, a deadlock exists in $W$ if and only if $W$ contains a
directed cycle.
\end{fact}

\begin{figure}[ht]
\centerline{\epsfbox{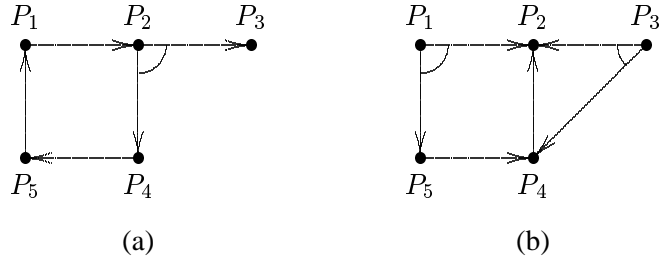}}
\caption{Two wait-for graphs in the AND model\label{graphand}.}
\end{figure}

In Figure~\ref{graphand}, we show two wait-for graphs in the AND model.
Circular arcs joining edges directed away from vertices are meant to indicate
that the AND model is being used. By Fact~\ref{andsuff}, there is deadlock in
the $W$ of Figure~\ref{graphand}(a), but not in Figure~\ref{graphand}(b).

The\inxx{deadlock detection, OR model}
case of the OR model is more subtle, and it is instructive to start by
realizing that the presence of a directed cycle in $W$ is no longer sufficient
for the existence of deadlocks. Clearly, so long as a directed path exists
in $W$ from every process to at least one sink, then no deadlock exists in
$W$ even though a directed cycle may be present. Formalizing this notion
requires that we consider the definition of a
{\em knot\/}\inxx{directed graph, knot}
in $W$.

For $P_i\in{\cal P}$, let ${\cal T}_i\subseteq{\cal P}$ be the set of vertices
that can be reached from $P_i$ through a directed path in $W$. This set
includes $P_i$ itself, and is known as the
{\em reachability set\/}\inxx{directed graph, reachability set}
of $P_i$ \cite{bm76}.
We say that a subset of vertices ${\cal S}\subseteq{\cal P}$ is a knot in $W$
if and only if ${\cal S}$ has at least two vertices and,
for all $P_i\in{\cal S}$, ${\cal T}_i={\cal S}$. By definition,
then, no member of a knot has a sink in its reachability set, which
characterizes the presence of a knot in $W$ as the sufficient condition we have
sought under the OR model. As it turns out, in fact, this condition is also
necessary, being stronger than the necessary condition established by
Fact~\ref{cycle}.

\begin{thm}\label{orsuff}
{\rm\cite{h72}}
In the OR model, a deadlock exists in $W$ if and only if $W$ contains a knot.
\end{thm}

\begin{figure}[ht]
\centerline{\epsfbox{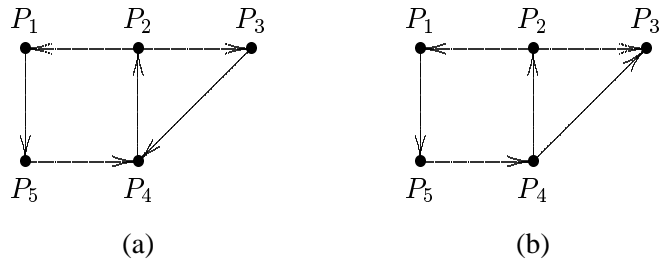}}
\caption{Two wait-for graphs in the OR model\label{graphor}.}
\end{figure}

The wait-for graphs of Figure~\ref{graphor} are for the OR model. A knot is
present in Figure~\ref{graphor}(a) (involving all vertices), but not in
Figure~\ref{graphor}(b). Thence, by Theorem~\ref{orsuff}, there is deadlock in
part (a) of the figure but not in part (b), in which $P_3$ is a sink that can
be reached from all other processes.

In order to identify sufficient conditions on $W$ that account for the existence
of deadlocks in the remaining deadlock models,
we must consider $W$ in a more explicit conjunction with the
deadlock model than we have done so far.
Let\inxx{deadlock detection, $x$-out-of-$y$ model}
us first consider the
$x$-out-of-$y$ model, and suppose that a subset of vertices
${\cal S}\subseteq{\cal P}$ can be identified having the property that, for
all $P_i\in{\cal S}$, $\vert{\cal O}_i\cap{\cal S}\vert>y_i-x_i$. Under these
circumstances, it is clear that no member of ${\cal S}$ can ever receive the
number of grant messages it requires, because at least one of such messages
would necessarily have to originate from within ${\cal S}$. In this paper,
we let a set such as ${\cal S}$ be called a
{\em $(y-x)$-knot}\inxx{directed graph, $(y-x)$-knot},
whose existence
in $W$ can also be shown to be necessary for deadlocks to exist. As in the
case of the OR model, this condition is stronger than the necessary condition
of Fact~\ref{cycle}.

\begin{thm}\label{xysuff}
{\rm\cite{ks94}}
In the $x$-out-of-$y$ model, a deadlock exists in $W$ if and only if $W$
contains a $(y-x)$-knot.
\end{thm}

\begin{figure}[ht]
\centerline{\epsfbox{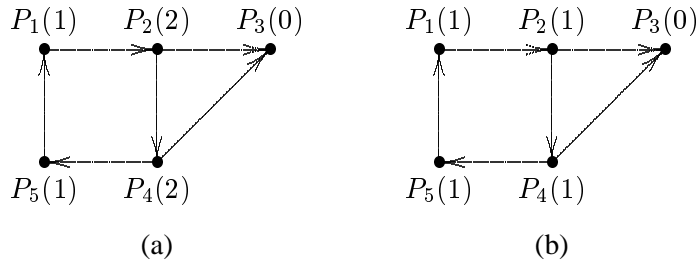}}
\caption{Two wait-for graphs in the $x$-out-of-$y$ model\label{graphxy}.}
\end{figure}

An illustration is given in Figure~\ref{graphxy}, with an integer in parentheses
next to the identification of each vertex to indicate its $x$ value. A
$(y-x)$-knot appears in Figure~\ref{graphxy}(a), but not in
Figure~\ref{graphxy}(b). The $(y-x)$-knot of Figure~\ref{graphxy}(a) involves
the vertices of the square. By Theorem~\ref{xysuff}, there is deadlock in the
$W$ of part (a) of the figure, but not in that of part (b). Note that $P_3$ is
a sink reachable from all vertices in both graphs, but this is to no avail in
the graph of part (a).

We\inxx{deadlock detection, AND-OR model}\inxx{deadlock detection, disjunctive $x$-out-of-$y$ model}
now turn to a discussion of sufficient conditions for deadlocks to exist in
$W$ under the AND-OR model. As we discussed in Section~\ref{models}, the
AND-OR model and the disjunctive $x$-out-of-$y$ model are equivalent to each
other, and for this reason the conditions that we come to identify as
sufficient under the AND-OR model will also be sufficient under the disjunctive
$x$-out-of-$y$ model if only we perform the transformation described in
Section~\ref{models}.

\begin{figure}[ht]
\centerline{\epsfbox{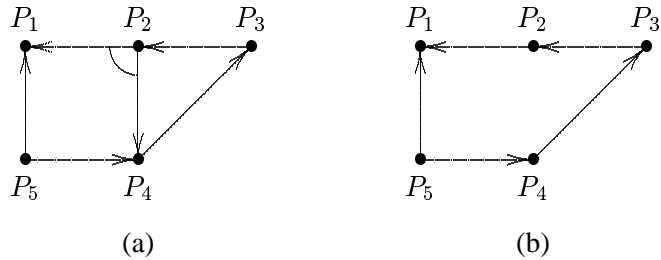}}
\caption{$W$ and one of its b-subgraphs\label{bsubg}.}
\end{figure}

Our starting point is the following definition. Consider a subgraph $W'$ of $W$
having vertex set ${\cal P}$, and for process $P_i$ let
${\cal O}'_i\subseteq{\cal P}$ be the set of vertices towards which directed
edges from $P_i$ exist in $W'$. In addition, for process $P_i$ let
${\cal O}'_i$ be such that
${\cal O}'_i\cap{\cal O}_i^1,\ldots,{\cal O}'_i\cap{\cal O}_i^{t_i}$
all have at least one member. We call such a subgraph a
{\em b-subgraph\/}\inxx{directed graph, b-subgraph}
of $W$, where
the ``b'' is intended to convey the notion that each directed edge in $W'$
relates to a ``bundle'' of directed edges stemming from the same vertex in $W$
\cite{bb98}.
An illustration is given in Figure~\ref{bsubg} of a wait-for graph in part (a)
and one of its b-subgraphs in part (b). Circular arcs around vertices in
Figure~\ref{bsubg}(a) indicate the ``AND'' groupings of neighbors that
constitute vertices' waits.

Intuitively, a b-subgraph of $W$ represents one of the various ``OR''
possibilities that are summarized in $W$ under the AND-OR model, provided that
we consider such possibilities ``globally,'' i.e., over all processes. As it
turns out, the existence of a knot in at least one of the b-subgraphs of $W$
is necessary and sufficient for a deadlock to exist in $W$. The knot that in
this case exists in that b-subgraph is called a
{\em b-knot\/}\inxx{directed graph, b-knot}
in $W$ \cite{bb98}.

\begin{thm}\label{andorsuff}
{\rm\cite{bb98}}
In the AND-OR model, a deadlock exists in $W$ if and only if $W$ contains a
b-knot.
\end{thm}

\begin{figure}[ht]
\centerline{\epsfbox{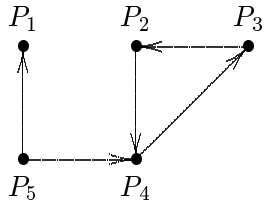}}
\caption{A b-subgraph with a knot\label{bknot}.}
\end{figure}

We show in Figure~\ref{bknot} another of the b-subgraphs of the wait-for graph
$W$ of Figure~\ref{bsubg}(a). This b-subgraph has a knot spanning the processes
in the triangle, which by Theorem~\ref{andorsuff} characterizes deadlock. In
fact, in $W$ it is easy to see that $P_2$ requires a relieve signal not only
from $P_1$ (this one must come eventually) but also from $P_4$ (which will
never come).

\section[Partially ordered sets and deadlock prevention]{Partially ordered sets and\\ deadlock prevention}\label{prevent}

In\inxx{deadlock prevention}
the remaining sections (Sections~\ref{prevent} through~\ref{color}), we
address the AND model exclusively. For this model, by Facts~\ref{cycle}
and~\ref{andsuff} we know that the existence of a directed cycle in the wait-for
graph $W$ is necessary and sufficient for a deadlock to exist. The fact that
this condition is necessary, in particular, allows us to look for design
strategies that prevent the occurrence of deadlocks beforehand by precluding
the appearance of directed cycles in $W$.

Of course, we also know from Fact~\ref{cycle} that directed cycles in $W$ are
necessary for a deadlock to exist regardless of the deadlock model. However,
for deadlock models other than the AND model, we have seen in
Section~\ref{detect} that there may exist a directed cycle in $W$ without the
corresponding existence of a deadlock. As a matter of fact, we have seen that
structures in $W$ much more complicated than directed cycles are necessary
for deadlocks to exist. Preventing the occurrence of cycles in those other
models is then too restrictive, while preventing the occurrence of the more
general structures appears to be too complicated. That is why our treatment of
deadlock prevention is henceforth restricted to the AND model.

Resource-sharing problems for the AND model are often referred to as the
{\em dining\/}\inxx{dining philosophers problem}
or {\em drinking\inxx{drinking philosophers problem}
philosophers problem\/} \cite{cm84,d68},
depending, respectively, on whether
every process $P_i$ always requests access to all the resources in ${\cal R}_i$
or not. In the remainder of this section, we discuss two prevention strategies
for such problems. Both strategies are based on the use of a
partially\inxx{partially ordered set}
ordered set (a poset), in the first case to establish dynamic priorities among
processes, in the second to establish a static global order for resource
request.

\subsection{Ordering the processes}\label{procorder}

Consider\inxx{deadlock prevention, ordering the processes}
the graph $G$ that represents the sharing of resources among processes,
and let $\omega$ be an
acyclic\inxx{undirected graph, acyclic orientation}
orientation of its edges. That is, $\omega$
assigns to each edge in ${\cal C}$ (the edge set of $G$) a direction in such a
way that no directed cycle is formed. This orientation establishes a partial
order on the set ${\cal P}$ of $G$'s vertices, so $G$ oriented by $\omega$ can
be regarded as a poset.

This poset is dynamic, in the sense that the acyclic orientation changes over
time, and can be used to establish priorities for processes that are adjacent
in $G$ to use shared resources when there is conflict. More specifically,
consider a resource-sharing computation that does the following. A process
sends requests for all resources that it needs, and must, upon receiving a
request, decide whether to grant access to the resource immediately or to do
it later. What the process does is to check whether the edge between itself
and the requesting process is oriented outwards by $\omega$. In the affirmative
case, it grants access to the resource either immediately or upon finishing to
use it (if this is the case). In the negative case, it may either grant access
(if it does not need the resource presently) or delay the granting until after
it has acquired all the resources it needs and used them. Whenever a process
finishes using a group of resources, it causes all edges presently oriented
towards itself to be oriented outward, thereby changing the acyclic orientation
of $G$ locally. These reversals of orientation constitute priority reversals
between the processes involved.

We see, then, that an acyclic orientation establishes a priority for resource
usage between every two neighbors in $G$, and that this priority is reversed
back and forth between them as they succeed in using the resources they need.
The crux of this mechanism is the simple property that the local changes a
process causes to the acyclic orientation always maintain its acyclicity,
and therefore its poset nature. If $\omega'$ is the acyclic orientation that
results from the application of such a local change, then we have the following.

\begin{lem}\label{acyclic1}
{\rm\cite{cm84}}
If $\omega$ is acyclic, then $\omega'$ is acyclic.
\end{lem}

\begin{figure}[ht]
\centerline{\epsfbox{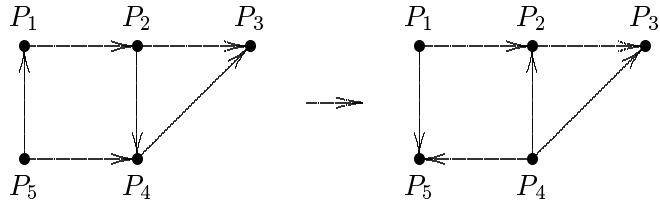}}
\caption{Edge reversal on acyclic orientations\label{ereverse}.}
\end{figure}

Note that Lemma~\ref{acyclic1} holds even if $\omega'$ results from local
changes applied to $\omega$ by more than one process concurrently. We show such
a pair of orientations in Figure~\ref{ereverse}, where the processes that do the
reversal are $P_1$ and $P_4$. In the dining-philosopher variant of the
resource-sharing computation, such a group of processes does necessarily
constitute an
{\em independent\inxx{undirected graph, independent set}
set\/} of $G$ (a set whose members are all
nonneighbors) \cite{bm76}.

The acyclicity of the changing orientation of $G$ is crucial in guaranteeing
that deadlocks never occur. To see this, consider the subgraph $W'$ of the
wait-for graph $W$ that corresponds to processes not being able to grant access
to resources immediately or after the (finite) time during which resources are
in use. According to the computation we outlined above, this happens when a
process, say $P_i$, receives a request from a neighbor $P_j$ but cannot grant it
immediately because it too needs the resource in question and furthermore holds
priority over $P_j$. Clearly, in both $W'$ and the acyclic orientation $\omega$
of $G$ that gives priorities, the edge between $P_i$ and $P_j$ is oriented from
$P_j$ to $P_i$. In other words, $W'$ is always a subgraph of $G$ oriented by
$\omega$, and by Lemma~\ref{acyclic1} never contains a directed cycle. Because
the edges that $W$ has in excess of $W'$ are all directed toward sinks, $W$ is
acyclic as well, which by Fact~\ref{cycle} implies the absence of deadlocks. If
we refer to computations such as the one we described as
{\em edge-reversal\inxx{resource-sharing computation, edge-reversal}
computations\/} \cite{b86}, then we have the following.

\begin{thm}\label{nodeadlock1}
{\rm\cite{cm84}}
Every edge-reversal computation is deadlock-free.
\end{thm}

Not only do the orientations of $G$ ensure the absence of deadlocks, but they
can be easily seen to ensure liveness guarantees as well. Owing to the
relationship of the computations we are considering to the dining-philosopher
paradigm, such guarantees are referred to as the absence of
{\em starvation}\inxx{resource sharing, starvation}.
That no starvation ever occurs comes also from the absence of directed cycles
in $W$: As the orientations of edges are reversed and $W$ evolves, the
farthest sinks for which a process is ultimately
waiting come ever closer to it, until
it too becomes a sink eventually and its wait ceases.

\begin{thm}\label{nostarvation1}
{\rm\cite{cm84}}
Every edge-reversal computation is starvation-free, and the worst-case wait a
process must undergo is $O(n)$.
\end{thm}

We note that, in Theorem~\ref{nostarvation1}, the wait of a process is measured
as the length of ``causal chains'' in the sending of grant messages, as is
customary in the field of asynchronous distributed algorithms \cite{b96}.

\subsection{Ordering the resources}\label{resorder}

\begin{figure}[ht]
\centerline{\epsfbox{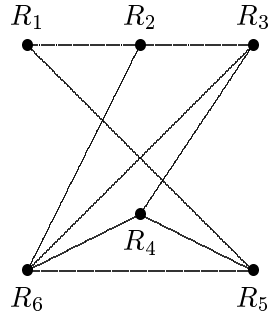}}
\caption{The graph $H$ for Example~\ref{basic}\label{graphh}.}
\end{figure}

The\inxx{deadlock prevention, ordering the resources}
graph $G$ that underlies all our resource-sharing computations has one
vertex per process and undirected edges connecting any two processes with the
potential to share at least one resource. The undirected graph we introduce now
and use throughout the end of the section is, by contrast, built on resources
for vertices, and has undirected edges connecting pairs of resources that are
potentially used in conjunction with each other by at least one process. This
graph is denoted by $H=({\cal R},{\cal E})$, where ${\cal E}$ contains an edge
between $R_p$ and $R_q$ if and only if ${\cal P}_{pq}\neq\emptyset$. $H$ is a
connected graph (because $G$ is connected) and contains a clique on ${\cal R}_i$
for $1\le i\le n$. The graph $H$ for Example~\ref{basic} is shown in
Figure~\ref{graphh}.

\begin{figure}[ht]
\centerline{\epsfbox{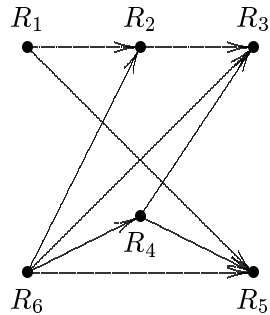}}
\caption{The graph $H$ for Example~\ref{basic} oriented acyclically
\label{digraphh}.}
\end{figure}

Our interest in graph $H$ comes from the possibility of constructing a poset on
its vertices by orienting its edges acyclically, similarly to what we did
previously on $G$. More specifically, let $\varphi$ be an acyclic orientation
of $H$, and for $R_p,R_q\in{\cal R}$ say that {\em $R_p$ precedes $R_q$} if and
only if a directed path exists from $R_p$ to $R_q$ in $H$ oriented by $\varphi$.
One acyclic orientation $\varphi$ for the graph of Figure~\ref{graphh} is given
in Figure~\ref{digraphh}. Note that the resources in ${\cal R}_i$ for any
$P_i\in{\cal P}$ are necessarily totally ordered by the ``precedes'' relation.

Now consider the following resource-sharing computation. When a process needs
access to a group of shared resources, it sends requests to the neighbors with
which it shares those resources according to the partial order implied by
$\varphi$. The rule to be followed is simple: A process only sends requests for
a resource $R_p$ after all grants have been received for the resources that it
needs and that precede $R_p$. The sending of grant responses to requests for a
particular resource $R_p$ is regulated by an $O(\vert{\cal P}_p\vert)$-time
distributed procedure on the vertices belonging to the clique in $G$ that
corresponds to that resource. This procedure for the acquisition of a single
resource must itself be deadlock- and starvation-free \cite{b96,wl93}.

Because $\varphi$ is acyclic, the evolving wait-for graph $W$ can never contain
a directed cycle, so by Fact~\ref{cycle} no deadlocks ever occur. The absence
of directed cycles in $W$ comes from the fact that such a cycle would imply a
``hold-and-wait'' cyclic arrangement of the processes, which is precluded by the
acyclicity of $\varphi$. We call these resource-sharing computations
{\em acquisition-order\inxx{resource-sharing computation, acquisition-order}
computations}, for which the following holds.

\begin{thm}\label{nodeadlock2}
{\rm\cite{l81}}
Every acquisition-order computation is deadlock-free.
\end{thm}

In addition to the safety guarantee given by Theorem~\ref{nodeadlock2}, and
similarly to the case of edge-reversal computations, for acquisition-order
computations it is also the case that liveness guarantees can be given.
In this case, however, liveness does not come from the shortening distance to
sinks in evolving acyclic orientations, but rather from the fact that directed
distances as given by an acyclic orientation are always bounded.

In order to be more specific regarding the liveness of acquisition-order
computations, let us consider a
{\em coloring\/}\inxx{undirected graph, coloring}
of the vertices of $H$. Such a
coloring is an assignment of colors (natural numbers) to vertices in such a way
that neighbors in $H$ get different colors. If $H$ can be colored with $c$
colors for some $c>0$, then we say that it is
{\em $c$-colorable\/}\inxx{undirected graph, $c$-colorable}
\cite{bm76}.

\begin{lem}\label{coloring}
{\rm\cite{d79}}
If $H$ is $c$-colorable, then there exists an acyclic orientation of $H$
according to which the longest directed path in $H$ has no more than $c-1$
edges.
\end{lem}

If process $P_i$ is the only one to be requesting resources in
an acquisition-order computation, then it waits for resources no longer than
is implied by the size of ${\cal R}_i$, the vertex set of a clique in $H$. So
its wait is given at most by the longest directed path in $H$ according to the
acyclic orientation $\varphi$ fixed beforehand. If $H$ is known to be
$c$-colorable, then by Lemma~\ref{coloring} $P_i$'s wait is bounded from
above by $c$. When other processes are also requesting resources, then let
$h$ be the maximum $\vert{\cal P}_p\vert$ for $1\le p\le m$, that is, the
maximum number of processes that may use a resource (this is the size of a
clique in $G$). We have the following.

\begin{thm}\label{nostarvation2}
{\rm\cite{l81}}
Every acquisition-order computation is starvation-free, and, if $H$ is
$c$-colorable, then the worst-case wait a process must undergo is $O(ch^c)$.
\end{thm}

In Figure~\ref{graphh}, the assignment of color $0$ to $R_3$ and $R_5$,
color $1$ to $R_2$ and $R_4$, and color $2$ to $R_1$ and $R_6$, makes the
graph $3$-colorable. One of the acyclic orientations complying with
Lemma~\ref{coloring} is the one shown in Figure~\ref{digraphh}.

Note, in all this discussion,
that it must be known beforehand that $H$ is $c$-colorable so
that $\varphi$ can be built. Also, by Theorem~\ref{nostarvation2}, it is to
one's advantage to seek as low a value of $c$ as can be efficiently found.
Seeking the optimal value of $c$ is equivalent to computing the graph's
{\em chromatic\inxx{undirected graph, chromatic number}
number\/} (the minimum number of colors with which the graph can
be colored) \cite{bm76}, and constitutes an NP-hard problem
\cite{gj79}.

\section{The graph abacus}\label{abacus}

In\inxx{graph abacus}\inxx{deadlock prevention, nonuniform access rates}
this section, we return to the edge-reversal computations discussed in
Section~\ref{procorder} and consider a generalization thereof in the special
context of high demand for resources by the processes. Such
a heavy-load situation occurs when, in the resource-sharing computation,
processes continually require access to all the shared resources they may have
access to, and endlessly go through an acquire-release cycle. Situations such
as this bring to the fore interesting
issues (some of which will be discussed in Section~\ref{color}) that are
relevant not only for the remainder of this section but also in the context of
our discussion in Section~\ref{procorder},
in which we addressed edge-reversal computations.

While it is conceivable that, in normal situations, such computations may
still be deadlock-free even if the corresponding orientations of $G$ have
cycles, the same cannot happen under heavy loads. This is so because what
those orientations do is to provide priority. In a light-load regime, a
cyclic dependency in the priority scheme may go unnoticed if the pattern of
resource demand by the processes happens never to cause a directed cycle in
$W$. Under heavy loads, on the other hand, the acyclicity of $G$'s orientations
is strictly necessary.

The generalization we consider is the following.
Associated with each process $P_i$ is an integer $r_i>0$ to be used to control
the dynamic evolution of priorities as given by the succession of acyclic
orientations of $G$. These numbers are to be used in such a way that, as
the computation progresses, and for any two neighbors
$P_i$ and $P_j$ in $G$, the ratio of the number of times $P_i$ has priority
over $P_j$ to the number of times $P_j$ has priority over $P_i$
``converges'' to $r_j/r_i$ in the long run \cite{bbf96,f94}.
The special case of Section~\ref{procorder} is obtained by setting $r_i$ to the
same number for all $P_i\in{\cal P}$. In that case, neighbors always have
alternating priorities and the ratio is therefore $1$.

We refer to computations with this generalized control of priorities as
{\em bead-reversal\inxx{resource-sharing computation, bead-reversal}
computations}, in allusion to the following implementation,
which views $G$'s edges as the rods along which the beads of a generalized
abacus (a graph abacus) are slid back and forth. For $(P_i,P_j)\in{\cal C}$,
let $e_{ij}$ beads be associated with edge $(P_i,P_j)$. In order for $P_i$ to
have priority over $P_j$, there has to exist at least $r_i$
beads on $P_i$'s side of the edge and strictly less than $r_j$ on $P_j$'s side.
When this is the case, the change in
priority is performed by moving $r_i$ of those beads towards $P_j$.

In an bead-reversal computation, the rule for process $P_i$ is the
following. Upon terminating its use of the shared resources, send $r_i$ beads
to the other end of every edge on which at least $r_i$ beads are on $P_i$'s
side. Under the assumption of heavy loads, this must be the case for all edges
adjacent to $P_i$, because under these circumstances processes can only access
resources when they have priority over all of their neighbors.

Just as with edge-reversal computations, it is possible to associate an
orientation of $G$'s edges to the priority scheme of bead-reversal computations.
For such, an edge is oriented towards $P_i$
if and only if there are at least $r_i$ beads on $P_i$'s side of the edge.
In order to preserve the syntactic constraints that an edge must be amenable to
orientation in any of the two possible directions, and that it has to be
oriented in exactly one direction at any time, we must clearly have
\[\max\{r_i,r_j\}\le e_{ij}\le r_i+r_j-1.\]

But it is possible to obtain a precise value for $e_{ij}$ within this range,
and also to come up with a criterion for an initial distribution of the beads
along the edges of $G$ in such a way as to provide the desired safety and
liveness guarantees. Safety is in this case associated with the acyclicity of
the orientations of $G$ as they change, while liveness refers to achieving
the desired ratios. As in the case of Section~\ref{procorder}, we aim at an
acyclic wait-for graph $W$ (for deadlock-freedom, by Fact~\ref{cycle}).
As for liveness, since achieving the desired ratios already implies
starvation-freedom, what we aim at are computations for which those ratios are
achieved, henceforth called
{\em ratio-compliant}\inxx{resource-sharing computation, ratio-compliant}.

We begin with the subgraph $G_{ij}$ of $G$ having for vertices the neighbors
$P_i$ and $P_j$ in $G$, along with the single edge between them. In what
follows, $g_{ij}$ is the greatest common divisor of $r_i$ and $r_j$.

\begin{thm}\label{ijgraph}
{\rm\cite{bbf96}}
If $e_{ij}=r_i+r_j-g_{ij}$, then every bead-reversal computation on $G_{ij}$ is
deadlock-free and ratio-compliant.
\end{thm}

\begin{figure}[ht]
\centerline{\epsfbox{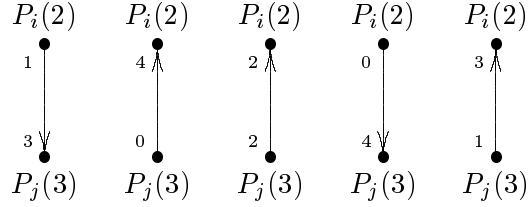}}
\caption{Bead reversal on $G_{ij}$\label{breverse}.}
\end{figure}

Theorem~\ref{ijgraph} makes no provisions as to the distribution of the $e_{ij}$
beads on the single edge of $G_{ij}$, and does as such hold for any of the
$(r_i+r_j)/g_{ij}$ possible distributions, as we see in Figure~\ref{breverse}.
In that figure, $r_i=2$ and $r_j=3$ (this is indicated in parentheses by the
vertices' identifications), and an evolution of bead placements is shown from
left to right. For each configuration, the number of beads on each of the
edge's ends is indicated by small numbers. The corresponding orientation of the
edge is also shown. Note that all five possible distributions of beads appear,
and that from the last one we return to the first.
When we consider the entirety of $G$,
however, the question of how to place the beads on $G$'s edges becomes crucial.

We begin the analysis of the general case by introducing some additional
notation and definitions. First, let $\Kappa$ denote the set of all the
simple cycles in $G$ (those with no repeated vertices). Membership of vertex
$P_i$ in $\kappa\in\Kappa$ is denoted by $P_i\in\kappa$, and membership of edge
$(P_i,P_j)$ in $\kappa\in\Kappa$ is denoted by $(P_i,P_j)\in\kappa$. Now, for
$\kappa\in\Kappa$, let $\kappa^+$ and $\kappa^-$ denote the two possible
traversal directions of $\kappa$, chosen arbitrarily. We use $a_{ij}^+$ to
denote the number of beads placed on edge $(P_i,P_j)$ on its far end as it is
traversed in the $\kappa^+$ direction, and $a_{ij}^-$ likewise for the
$\kappa^-$ direction. Obviously, at
all times we have $a_{ij}^++a_{ij}^-=e_{ij}$.

For $\kappa\in\Kappa$, let
\[\rho(\kappa)=\sum_{P_i\in\kappa}r_i\]
and
\[\sigma(\kappa)
=\max\left\{
\sum_{(P_i,P_j)\in\kappa}a_{ij}^+,\sum_{(P_i,P_j)\in\kappa}a_{ij}^-
\right\}.\]
According to these equations, $\rho(\kappa)$ is the sum of $r_i$ over all
vertices $P_i$ of $\kappa$, while $\sigma(\kappa)$ denotes the total number
of beads found on $\kappa$'s edges' far ends as $\kappa$ is traversed in the
$\kappa^+$ direction or along the $\kappa^-$ direction, whichever is greatest.
In addition, it is easy to see that both $\rho(\kappa)$ and $\sigma(\kappa)$
are time-invariant. We are now ready to state the counterpart of
Theorem~\ref{ijgraph} for $G$ as a whole.

\begin{thm}\label{entiregraph}
{\rm\cite{bbf96}}
If $e_{ij}=r_i+r_j-g_{ij}$ for all $(P_i,P_j)\in{\cal C}$, then every
bead-reversal computation is deadlock-free and ratio-compliant if and
only if $\sigma(\kappa)<\rho(\kappa)$ for all $\kappa\in\Kappa$.
\end{thm}

One interesting special case that can be used to further our insight into
the dynamics of bead-reversal computations is the case of graphs without
(undirected) cycles, that is, cases in which $G$ is a tree. In such cases,
$\Kappa=\emptyset$ and Theorem~\ref{entiregraph} becomes a simple generalization
(by quantification over all of ${\cal C}$) of Theorem~\ref{ijgraph}.

To finalize this section, we return to the wait-for graph $W$ to analyze its
acyclicity. As in the case of edge-reversal computations
(cf.\ Section~\ref{procorder}), $W$ is related to an oriented version of $G$,
as follows.
In bead-reversal computations, the orientation of edge $(P_i,P_j)$ is from
$P_i$ towards $P_j$ if and only if there are at least $r_j$ beads placed on the
$P_j$ end of the edge (and, necessarily, fewer than $r_i$ on $P_i$'s end).
Recalling, as in Section~\ref{procorder}, that $W$ is the graph the results when
processes cannot send grant messages, then the heavy-load assumption implies
that $W$ coincides with $G$ oriented as we just discussed. What this means is
that Theorem~\ref{entiregraph} is an indirect statement on the acyclicity of
$W$: If a directed cycle exists in $W$, then obviously for the corresponding
underlying $\kappa$ we have $\sigma(\kappa)\ge\rho(\kappa)$, which
characterizes an orientation of $G$ that is not acyclic either.

But the attentive reader will have noticed that violating the inequality of
Theorem~\ref{entiregraph} does not necessarily lead to a directed cycle in
$G$'s orientation (or in $W$). The significance of the theorem, however, is
that such a cycle is certain to be created at some time if the inequality is
violated. What the theorem does is to provide a criterion for the establishment
of initial conditions (bead placement) that is necessary even though at first
no cycle might be created otherwise.

\begin{figure}[ht]
\centerline{\epsfbox{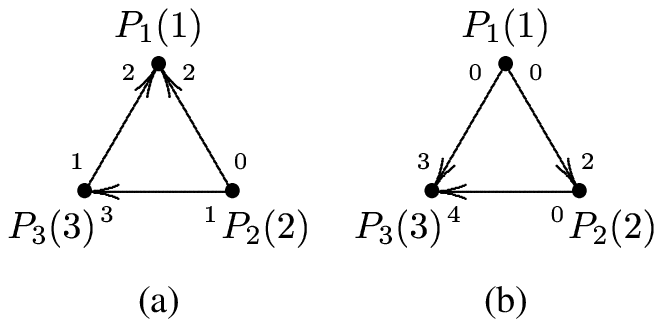}}
\caption{Two possible bead placements\label{beadplc}.}
\end{figure}

\begin{exmp}\label{2abaci}
Let $G$ be the complete graph on three vertices, and let $r_1=1$, $r_2=2$,
and $r_3=3$. Then $e_{12}=2$, $e_{13}=3$, and $e_{23}=4$. Employing the same
convention as in Figure~\ref{breverse}, and identifying the $\kappa^+$
direction of traversal with the clockwise direction for the single simple
cycle $\kappa$, we show in Figure~\ref{beadplc} two possibilities for
bead placement. The one in part (a) has $\sigma(\kappa)=5$, while
$\sigma(\kappa)=6$ for part (b), both values determined by the $\kappa^+$
direction. We have $\rho(\kappa)=6$ for this example,
so $\sigma(\kappa)<\rho(\kappa)$ in part (a), whereas
$\sigma(\kappa)=\rho(\kappa)$ in part (b). Although both orientations are
acyclic, the reader can check easily that the evolution of the bead placement
in Figure~\ref{beadplc}(b) will soon lead to a directed cycle, while for the
other acyclicity will be indefinitely preserved. This is, of course, in
accordance with Theorem~\ref{entiregraph}.
\end{exmp}

\section[Graph coloring and concurrency]{Graph coloring and\\ concurrency}\label{color}

From\inxx{resource sharing, concurrency}
a purely algorithmic perspective, heavy-load situations such as introduced
in the beginning of Section~\ref{abacus} provide a simpler means of implementing
edge-reversal and bead-reversal computations than the overall scheme discussed
in Section~\ref{comp}. In a heavy-load regime, the need for processes to
explicitly request and grant resources becomes moot, because the reversal
of priorities (edge orientation or beads) can be taken to signify that
permission is granted (or partially granted, in the case of bead reversals) to
access shared resources.

Given this simplification, the following is how an edge-reversal computation
can be implemented. Start with an acyclic orientation of $G$. A process computes
on shared resources when it is a sink, then reverses all edges adjacent to it
and waits to become a sink once again. Similarly, and following our anticipation
in Section~\ref{abacus}, a bead-reversal computation is also
simple, as follows. Start with a placement of the beads that not only leads
to an acyclic orientation of $G$ but also complies with the inequality
prescribed by Theorem~\ref{entiregraph}. A process $P_i$ computes when it is a
sink (has enough beads on all adjacent edges), then sends $r_i$ beads to each
of its neighbors in $G$. This is repeated until $P_i$ ceases being a sink, at
which time it waits to become a sink again.

Heavy-load situations also raise the question of how much concurrency, or
parallelism, there can be in the sharing of resources. While neighbors in $G$
are precluded from sharing resources concurrently, processes that are not
neighbors can do it, and how much of it they can do depends on the initial
conditions that are imposed on the computation (acyclic orientation of $G$ or
bead placement). In the remainder of this section, we discuss this issue of
concurrency for edge-reversal computations only, but a similar discussion can be
done for bead-reversal computations as well \cite{bbf96}.

The simplest means to carry out this concurrency analysis is to abandon the
asynchronous model of computation we have been assuming
(cf.\ Section~\ref{comp})
and to assume full synchrony instead. In the {\em fully synchronous\/}
(or simply
{\em synchronous\/})\inxx{synchronous model of distributed computing}
model of distributed computation \cite{b96},
processes are driven by
a common global clock that issues ticks represented by the integer $s\ge 0$.
At each tick, processes compute and send messages to their neighbors, which are
assumed to get those messages before the next tick comes by.

An edge-reversal computation under the synchronous model is an infinite
succession of acyclic orientations of $G$. If these orientations are
$\omega_0,\omega_1,\ldots,$ then, for $s>0$, $\omega_s$ is obtained from
$\omega_{s-1}$ by turning every sink in $\omega_{s-1}$ into a
{\em source\/}\inxx{directed graph, source}
(a vertex with all adjacent edges directed outward). The number of distinct
acyclic orientations of $G$ is finite, so the sequence
$\omega_0,\omega_1,\ldots$ does eventually become periodic, and from this point
on it contains an endless repetition of a number of orientations that we denote
by $p(\omega_0)$ (this notation is meant to emphasize that the acyclic
orientations that are repeated periodically are fully determined by $\omega_0$).
Let these $p(\omega_0)$ orientations be called the {\em periodic orientations\/}
from $\omega_0$.

\begin{lem}\label{period}
{\rm\cite{bg89}}
The number of times a process is a sink in the periodic orientations from
$\omega_0$ is the same for all processes.
\end{lem}

We let $m(\omega_0)$ denote the number asserted by Lemma~\ref{period}, and let
$m_i(s)$ denote the number of times process $P_i$ is a sink in the subsequence
$\omega_0,\ldots,\omega_{s-1}$.

Intuitively, it should be obvious that the amount of concurrency achieved from
the initial conditions given by $\omega_0$ depends chiefly on the periodic
repetition that is eventually reached. In order to make this more formal,
let ${\it Conc}(\omega_0)$ denote this amount of concurrency, and define it as
\[{\it Conc}(\omega_0)
=\lim_{s\to\infty}{{1}\over{sn}}\sum_{P_i\in{\cal P}}m_i(s).\]
That is, we let the concurrency from $\omega_0$ be the average, taken over
time and over the number of processes, of the total number of sinks in the
sequence $\omega_0,\ldots,\omega_{s-1}$ as $s\to\infty$ (the existence of this
limit, which is implicitly assumed by the definition of ${\it Conc}(\omega_0)$,
is only established in what follows, so the definition is a little abusive for
the sake of notational simplicity).

\begin{thm}\label{concmp}
{\rm\cite{bg89}}
${\it Conc}(\omega_0)=m(\omega_0)/p(\omega_0)$.
\end{thm}

Theorem~\ref{concmp} characterizes concurrency in a way that emphasizes the
dynamics of edge-reversal computations under heavy loads. But the question that
still remains is whether a characterization of concurrency exists that does
not depend on the dynamics to be computed, but rather follows from the structure
of $G$ as oriented by $\omega_0$.

This question can be answered affirmatively, and for that we consider once again
the set $\Kappa$ of all simple cycles in $G$. For $\kappa\in\Kappa$, we let
$c^+(\kappa,\omega_0)$ be the number of edges in $\kappa$ that are oriented
by $\omega_0$ in one of the two possible traversal directions of $\kappa$.
Likewise for $c^-(\kappa,\omega_0)$ in the other direction. The number of
vertices in $\kappa$ is denoted by $\vert\kappa\vert$.

\begin{thm}\label{concgraph}
{\rm\cite{bg89}}
If $G$ is a tree, then ${\it Conc}(\omega_0)=1/2$. Otherwise, then
\[{\it Conc}(\omega_0)
=\min_{\kappa\in\Kappa}
{{\min\{c^+(\kappa,\omega_0),c^-(\kappa,\omega_0)\}}\over{\vert\kappa\vert}}.\]
\end{thm}

Except for the case of trees, by Theorems~\ref{concmp} and~\ref{concgraph} we
know that the amount of concurrency of an edge-reversal computation is entirely
dependent upon $\omega_0$, the initial acyclic orientation. The problem of
determining the $\omega_0$ that maximizes concurrency is, however,
NP-hard, so an exact efficient procedure to do it is unlikely to exist
in general \cite{bg89}.

\begin{figure}[ht]
\centerline{\epsfbox{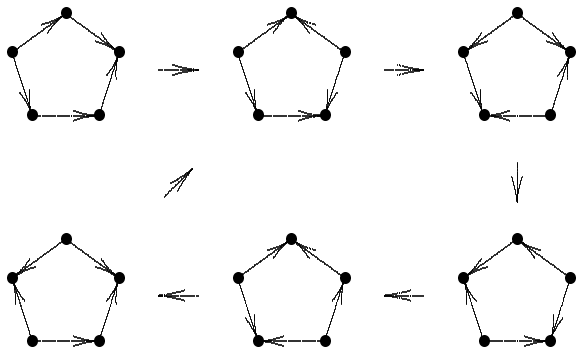}}
\caption{A heavy-load case of edge reversal\label{ser}.}
\end{figure}

\begin{exmp}\label{dpp}
When $G$ is a ring on five vertices, we have a representation of the original
dining philosophers problem \cite{d68}. For this case, consider the
sequence of acyclic orientations depicted in Figure~\ref{ser}, of which any one
can be taken to be $\omega_0$. We have $m(\omega_0)=2$, $p(\omega_0)=5$, and
${\it Conc}(\omega_0)=2/5$. This concurrency value follows from either
Theorem~\ref{concmp} or Theorem~\ref{concgraph}.
\end{exmp}

Another interesting facet of this concurrency issue is that it relates closely
to various forms of coloring the vertices of $G$. Consider, for example, the
{\em $k$-tuple\inxx{undirected graph, $k$-tuple coloring}
coloring\/} of the vertices of $G$ obtained as follows
\cite{s76}. Assign
$k$ distinct colors to each vertex in such a way that no two neighbors share
a color. This type of coloring generalizes the coloring discussed in
Section~\ref{resorder}, for which $k=1$. The minimum number of colors required
to provide $G$ with a $k$-tuple coloring is its
{\em $k$-chromatic\inxx{undirected graph, $k$-chromatic number}
number}.

In the context of edge-reversal computations, note that the choice of an initial
acyclic orientation $\omega_0$ implies, by Lemma~\ref{period}, that $G$ admits
an $m(\omega_0)$-tuple coloring requiring a total of $p(\omega_0)$ colors. If
these colors are natural numbers, then neighbors in $G$ get colors that are
``interleaved,''\inxx{undirected graph, interleaved colors}
in the following sense. For two neighbors $P_i$ and $P_j$,
let $c_i^1,\ldots,c_i^z$ and $c_j^1,\ldots,c_j^z$ be their colors, respectively,
with $z=m(\omega_0)$. Then either $c_i^1<c_j^1<\cdots<c_i^z<c_j^z$ or
$c_j^1<c_i^1<\cdots<c_j^z<c_i^z$.

So the question of maximizing concurrency is, by Theorem~\ref{concmp},
equivalent to the question of minimizing the ratio of the total number of
interleaved colors to the number of colors per vertex (this is the ratio
$p(\omega_0)/m(\omega_0)$) by choosing $\omega_0$ appropriately.
The optimal ratio thus obtained, denoted by
$\bar{\chi}(G)$, is called the
{\em interleaved\inxx{undirected graph, interleaved multichromatic number}
multichromatic\/} (or
{\em interleaved fractional chromatic\/}) {\em number of $G$} \cite{bg89}.
When the interleaving of
colors is not an issue, then what we have is the graph's
{\em multichromatic\/}\inxx{undirected graph, multichromatic number}
(or {\em fractional chromatic\/}) {\em number\/} \cite{su97}.

\begin{figure}[ht]
\centerline{\epsfbox{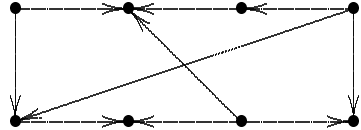}}
\caption{A graph $G$ for which $\chi^*(G)<\bar{\chi}(G)<\chi(G)$
\label{alldiff}.}
\end{figure}

Letting $\chi(G)$ denote the
chromatic number of $G$ and $\chi^*(G)$ its multichromatic number, we have
\[\chi^*(G)\le\bar{\chi}(G)\le\chi(G).\]
A graph $G$ is shown in Figure~\ref{alldiff} for which $\chi^*(G)=5/2$,
$\bar{\chi}(G)=8/3$, and $\chi(G)=3$, all distinct therefore. One of the
orientations that correspond to $\bar{\chi}(G)=8/3$ is the one shown in the
figure.

\section{Concluding remarks}\label{concl}

Distributed computations over shared resources are complex, asynchronous
computations. Performing such computations efficiently while offering a
minimal set of guarantees has been a challenge for several decades. At present,
though problems still persist, we have a clear understanding of several of the
issues involved and have in many ways met that challenge successfully.

Crucial to this understanding has been the use of precise modeling tools,
aiming primarily at clarifying the timing issues involved, as well as the
combinatorial structures that underlie most of concurrent computations. In this
paper, we have concentrated on the latter and outlined some of the most
prominent combinatorial concepts on which the design of resource-sharing
computations is based. These have included graph structures and posets useful
for handling the safety and liveness issues that appear in those computations,
and for understanding the questions related to concurrency.

\begin{acknowledgments}
The author is thankful to Mario Benevides and Felipe Fran\c ca for many fruitful
discussions on the topics of this paper.
\end{acknowledgments}


\input{barbosa.bbl}
\end{document}

%% file: def.tex
\newtheorem{defnt}{Definition}[chapter]
\newtheorem{lem}[defnt]{Lemma}
\newtheorem{thm}[defnt]{Theorem}

\newtheorem{fact}[defnt]{Fact}
\newtheorem{exmp}{Example}[chapter]